\documentclass[a4paper]{jpconf}
\usepackage{graphicx}
\usepackage{arydshln}

\begin{document}
\title{Supernova Early Warning in Daya Bay Reactor Neutrino Experiment}

\author{Hanyu Wei for the Daya Bay Collaboration}

\address{Department of Engineering Physics, Tsinghua University, Beijing, China}

\ead{weihy07@mails.tsinghua.edu.cn}

\begin{abstract}
Providing an early warning of a galactic supernova using neutrino signals is of importance in studying both supernova dynamics and neutrino physics. The Daya Bay reactor neutrino experiment, with a unique feature of multiple liquid scintillator detectors separated in space, is sensitive to the full energy spectrum of supernova burst electron-antineutrinos. By deploying 8 Antineutrino Detectors (ADs) in three different experimental halls, we obtain a more powerful and prompt rejection of muon spallation background than single-detector experiments. A dedicated supernova online trigger system embedded in the data acquisition system has been installed to allow the detection of a coincidence of neutrino signals within a 10-second window, thus providing a robust early warning of a supernova occurrence within the Milky Way.
\end{abstract}

\section{Motivation}
The Daya Bay reactor neutrino experiment is specifically designed for measuring the neutrino mixing angle $\theta_{13}$ with a sensitivity down to the 1\% level \cite{dyb}. However, the deployment of 8 electron-antineutrino detectors (8 ADs) in three different experimental halls (Daya Bay near site, Ling Ao near site, and Far site) motivates studies for a supernova online trigger without complicated reconstruction and offline analysis. The three experimental halls are more than 1 km apart from each other, which enables a powerful and prompt rejection of muon-induced and accidental backgrounds superior to that of a single-detector. In addition, a relatively low energy threshold of 0.7 MeV enhances the detection of the full energy spectrum of supernova burst neutrinos (SN$\nu$) since the energy spectrum may vary according to the supernova core collapse model. A supernova online trigger system is installed and in preparation to join the Supernova Early Warning System (an international organization abbreviated to SNEWS), providing the astronomical community with a prompt alert of the occurrence of a galactic core collapse event \cite{snewsurl} with the false alarm rate $<$ 1/month.

\section{Detection of electron-antineutrinos in Daya Bay}
\label{dayabay}
 The ADs in Daya Bay are designed to detect the $\bar{\nu}_e$ via inverse beta decay (IBD) interactions $\bar{\nu}_e + p \rightarrow n + e^+$. Each single AD has 22 tons of liquid scintillator (LS) and 20 tons of liquid scintillator doped with gadolinium (Gd-LS), giving a total target mass of $\sim$330 tons in 8 ADs. The coincidence of the prompt scintillation from the $e^+$ and the delayed gamma emission of the neutron capture provides a distinctive $\bar{\nu}_e$ signature against the background. The dominant backgrounds are accidentals, cosmogenically produced fast neutrons, $^9$Li/$^8$He decays and the neutrons from the retracted $^{241}$Am-$^{13}$C calibration source. The average delay of the gamma emission of the neutron capture is 28 $\mu$s for gadolinium and 200 $\mu$s for hydrogen. \cite{dayabay1, dayabay2}

\section{Neutrino emission from supernovae}
Supernova burst neutrinos (consisting of $\nu_e, \bar{\nu}_e, \nu_{\mu}, \bar{\nu}_{\mu}, \nu_{\tau}, \bar{\nu}_{\tau}$) play a role in the study of both supernova dynamics and neutrino physics, because

\begin{itemize}
\item $\sim$99\% of the stellar collapse gravitational binding energy is converted to neutrinos which arrive at the earth a few hours before the visual supernova explosion (SNe). So, it is believed that neutrino emission and interaction are a key diagnostic for the dynamics of core collapse and supernova explosion \cite{snevent};
\item Supernova burst neutrinos can serve as probes of neutrino properties, e.g. neutrino mixing, neutrino mass, neutrino lifetime, magnetic moment of neutrino, electric charge of neutrino, radiative decay of neutrino, etc. \cite{snvprobe} and also the mass hierarchy \cite{snvmasshierarchy} ;
\item Joint analysis with gravitational waves can provide deep insight into the core collapse of supernovae \cite{gravitational}.
\end{itemize}

The expected SN explosion rate is $\sim$0.01/year \cite{snrate} within kilo-parsec (kpc) distances and is around once per 50 years in the Milky Way. Within Mpc distances, the rate is $\sim$1/year \cite{snrate}, but the neutrino flux is much smaller. The SN$\nu$ energy spectrum within the first 10 seconds of the supernova exlosion \cite{energyspectra} for different flavor components implies the energy range of electron-antineutrinos is up to $\sim$60 MeV with average energy 12$\sim$15 MeV. The explosion timescale is $\sim$10 s with $\sim$98\% of the $\bar{\nu}_e$ luminosity emitted \cite{timingspectra}. This timing feature is exploited to form an online trigger for SN$\nu$ in all the experiments listed in Tab.~\ref{tab:SN_event}, where the main features are summarized. Based on the target mass, Tab.~\ref{tab:SN_event} shows that about 12 SN$\nu$ events in one AD and 100 events in all for 8-ADs are expected at Daya Bay and a SN$\nu$ event is defined as the detection of one neutrino from a single SN explosion. Even though other experiments may have higher expected SN$\nu$ events mainly due to the target mass, it is emphasized that the unique feature of Daya Bay in contrast is that it is not a single-detector. This paper explains this advantage and shows that the Daya Bay experiment is sensitive to all the 1987A-type (referring to the luminosity and average energy of $\bar{\nu}_e$) SNe in the Milky Way which can be seen in Fig.~\ref{fig:detectionprobability}.

\begin{center}
\begin{table}[h!]
\centering
\caption{\label{tab:SN_event}SN$\nu$ sensitive detectors and expected events for a SN at 10 kpc, emission of $5\times10^{52}$ erg in $\bar{\nu}_e$, average energy 12 MeV, compatible with SN1987A. \cite{snevent}}
\begin{tabular}{llllll}
\br
Detector & Type & Location & Mass[kt] & Events & Status \\
\mr
IceCube & Ice Cherenkov & South Pole & 0.6/OM & $10^6$ & Running \\
Super-K IV & Water & Japan & 32 & 7000 & Running \\
LVD & Scintillator & Italy & 1 & 300 & Running \\
KamLAND & Scintillator & Japan & 1 & 300 & Running \\
SNO+ & Scintillator & Canada & 1 & 300 & Commissioning 2013 \\
MiniBOONE & Scintillator & USA & 0.7 & 200 & Running \\
\textbf{Daya Bay} & \textbf{Scintillator} & \textbf{China} & \textbf{0.33} & \textbf{100} & \textbf{Running} \\
Borexino & Scintillator & Italy & 0.3 & 80 & Running \\
BST & Scintillator & Russia & 0.2 & 50 & Running \\
HALO & Lead & Canada & 0.079 & tens & Almost ready \\
ICARUS & Liquid argon & Italy & 0.6 & 200 & Running \\
\br
\end{tabular}
\end{table}
\end{center}

\section{Background sources and the supernova burst neutrino event}
The supernova online trigger system in Daya Bay is embedded in the Data Acquisition System (DAQ), online looking for increase in multi-AD signals in 10s-time-window and sending prompt alarms. According to this task, all the study of supernova online trigger is for online prompt trigger judgment and not so precise as offline analysis. The purpose of the background study on one hand is to have a good understanding of the background coincidences in multi-AD, thus allowing to set a precise false alarm rate threshold. The false alarm happens frequently as the detectable SN explosion to the earth is so rare and the selected events are always backgrounds. On the other hand, the background study contributes to the event selection criteria establishment which has to be simpler than that of the offline analysis so as to be prompt and not to bring much workload to DAQ. A data sample from Dec. 24, 2011 to Jul. 28, 2012 is used to train our online trigger algorithm to give the event selection criteria and study the backgrounds since no observation of SN$\nu$ was declared during the period of the data sample by all detectors including Daya Bay. In addition, the supernova burst neutrinos that undergo an IBD in the detector volume are simulated aiming to study the detection efficiency of SN$\nu$.

\subsection{Background sources}
\begin{figure}
\begin{center}
\includegraphics[width=8cm]{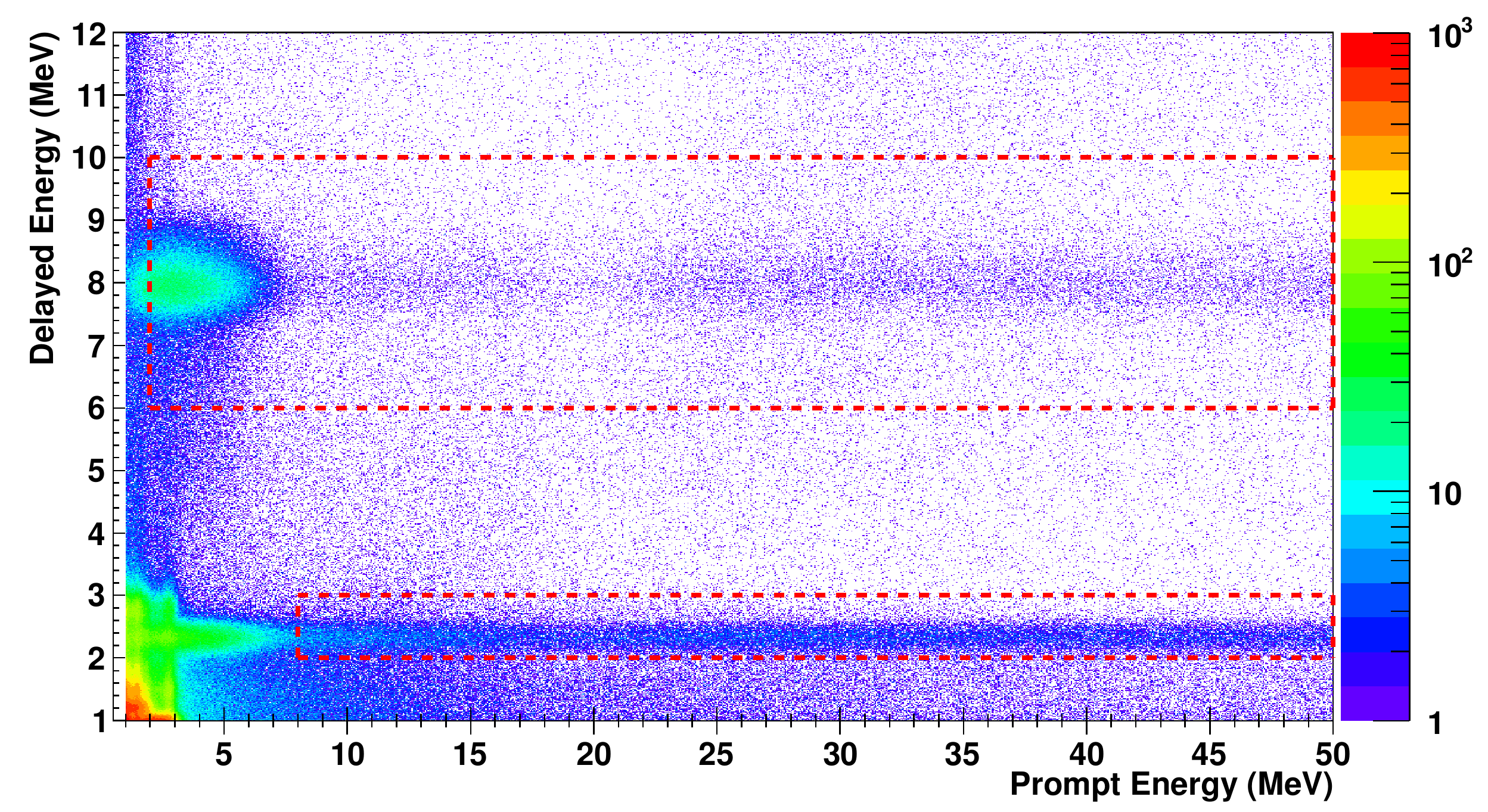}
\end{center}
\caption{Prompt signal energy vs. delayed signal energy 2-D plot for backgrounds. In the red box is the selection region for SN$\nu$, suggesting the prompt and delayed energy cut.}
\label{fig:Background2D}
\end{figure}

In the Daya Bay ADs (Section~\ref{dayabay}), referring to Fig.~\ref{fig:Background2D}, the delayed signal of an IBD event is either an 8 MeV $\gamma$ cascade from neutron capture on Gd, or a 2.2 MeV $\gamma$ from neutron capture on H. It is observed that the large amount of accidental backgrounds in the low energy region significantly affect the background event rate, therefore we set the online energy threshold at 2 MeV for the prompt signal associated with the 8 MeV $\gamma$ cascade and 8 MeV for that associated with single 2.2 MeV $\gamma$ in which case the majority of accidental backgrounds are removed. Using optimized selection criteria for SN$\nu$, the prompt vs. delayed signal energy plot is shown in the red box in Fig.~\ref{fig:Background2D}. Along the Y-axis, the Gd neutron capture peak is seen around 8 MeV and the hydrogen neutron capture peak is around 2.2 MeV where many fast neutrons are present in the high energy range along the X-axis and reactor neutrino signals are present below 10 MeV.

Notice that the data for trigger algorithm training are offline reconstructed while the supernova online trigger can only access the raw data. A simple but relatively effective reconstruction is applied online for real SN$\nu$ selection in which the average PMT gain and energy scale calibration constants are used for energy reconstruction and a charge-weighted method is used for prompt vertex reconstruction. The resulting online, measured single AD event rates are 0.019, 0.013 and 0.0013 Hz/AD at the Daya Bay near site, Ling Ao near site and far site, respectively.

\subsection{Supernova burst neutrino event}
Assuming that the spectrum of supernova burst neutrinos follows a quasithermal distribution \cite{snmodel}
$$
f_{\nu}(E)\propto E^{\alpha}e^{-(\alpha+1)E/E_{av}}
$$
where $E_{av}$ is the average energy and $\alpha$ a numerical parameter describing the amount of spectral pinching. The value $\alpha$ = 2.30 corresponds to a Fermi-Dirac distribution with zero chemical potential. In our simulation $\alpha$ is $>$ 2.30 and varies with the three main phases of the detectable supernova neutrino signals: prompt $\nu_e$ burst phase, accretion phase and cooling phase \cite{snevent}.

\begin{figure}[h!]
\centering
\includegraphics[width=5.3cm, height=3.1cm]{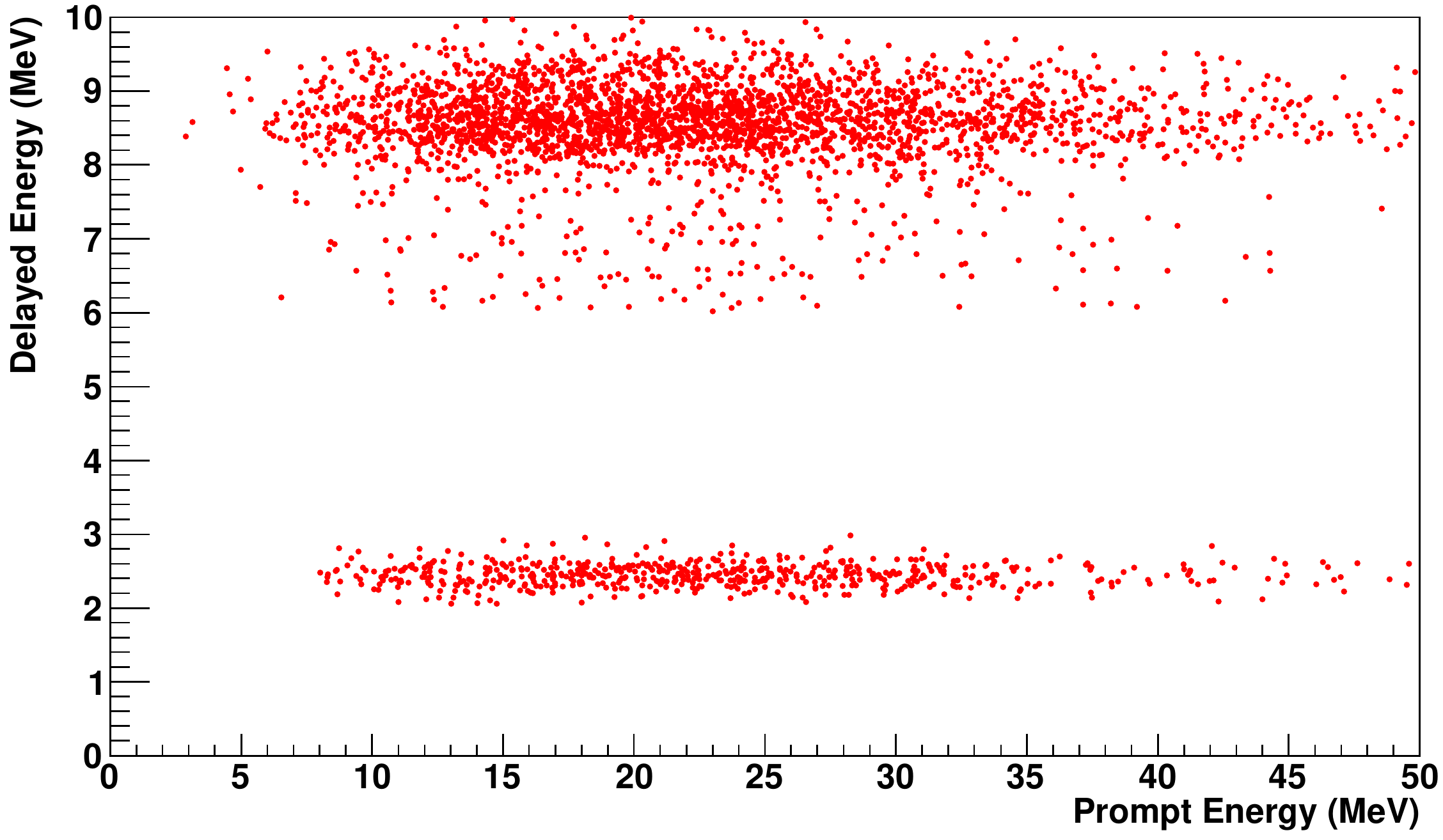}
\includegraphics[width=5.3cm, height=3.1cm]{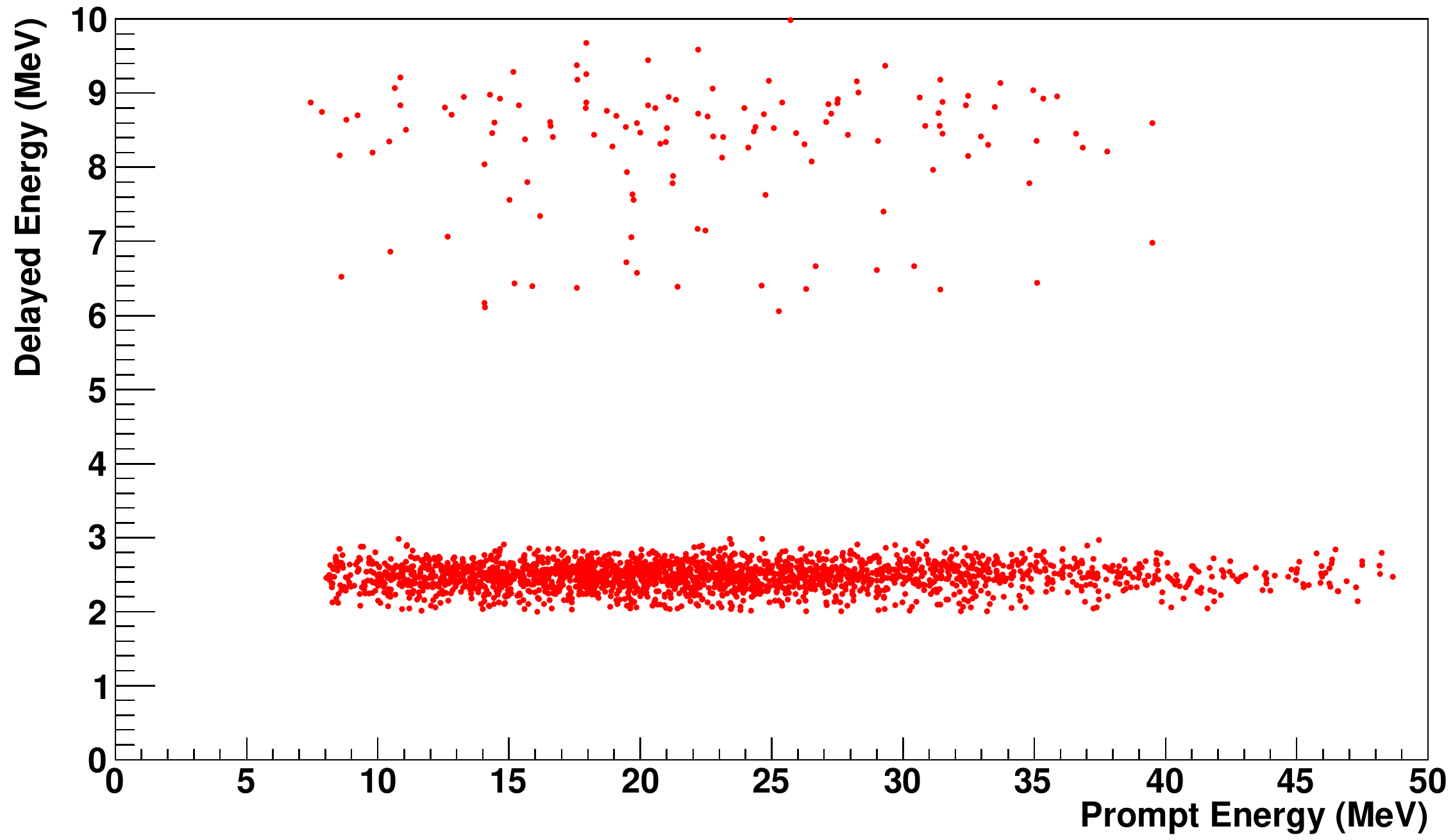}
\includegraphics[width=5.3cm, height=3.1cm]{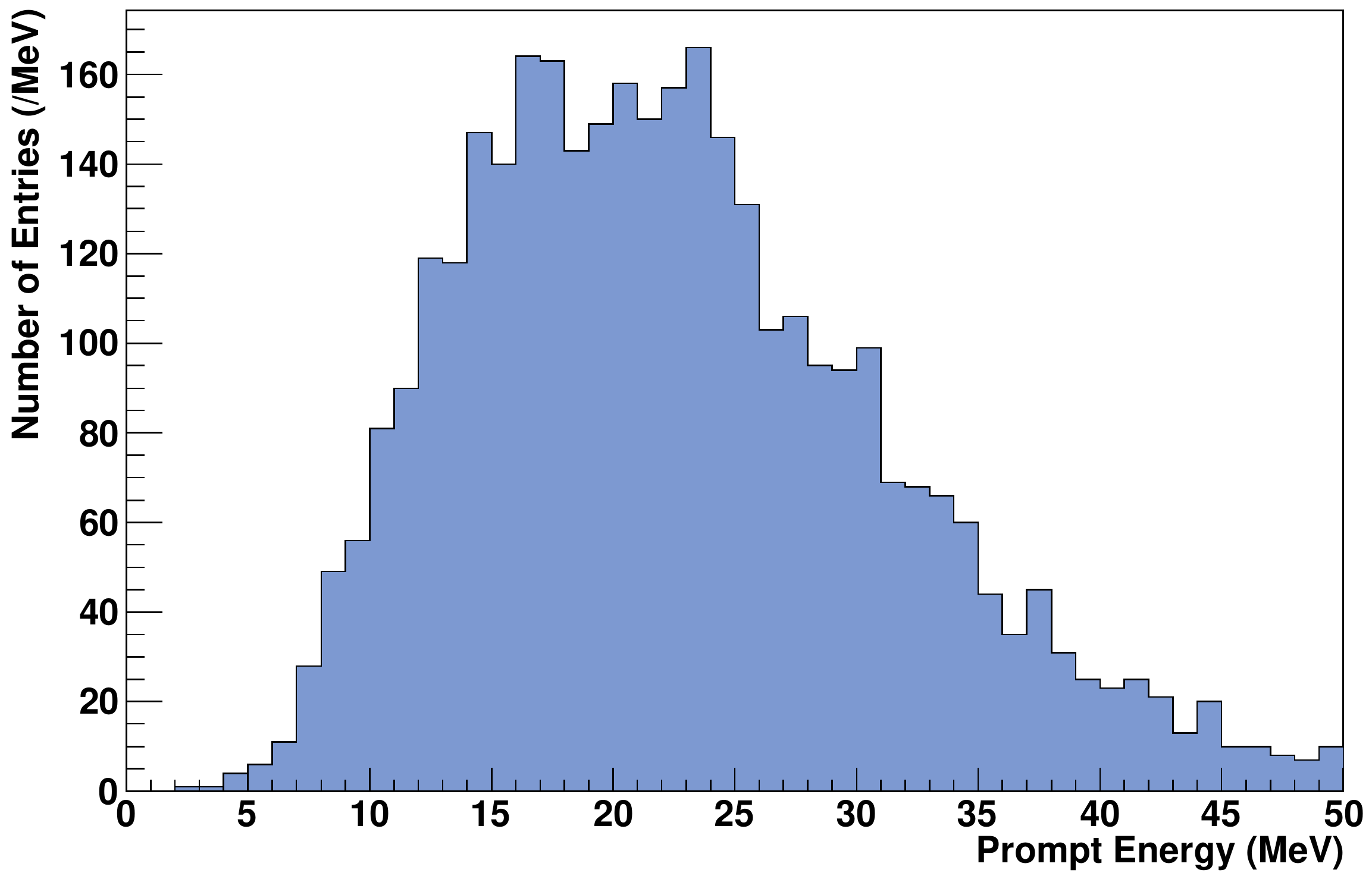}
\includegraphics[width=5.3cm, height=3.1cm]{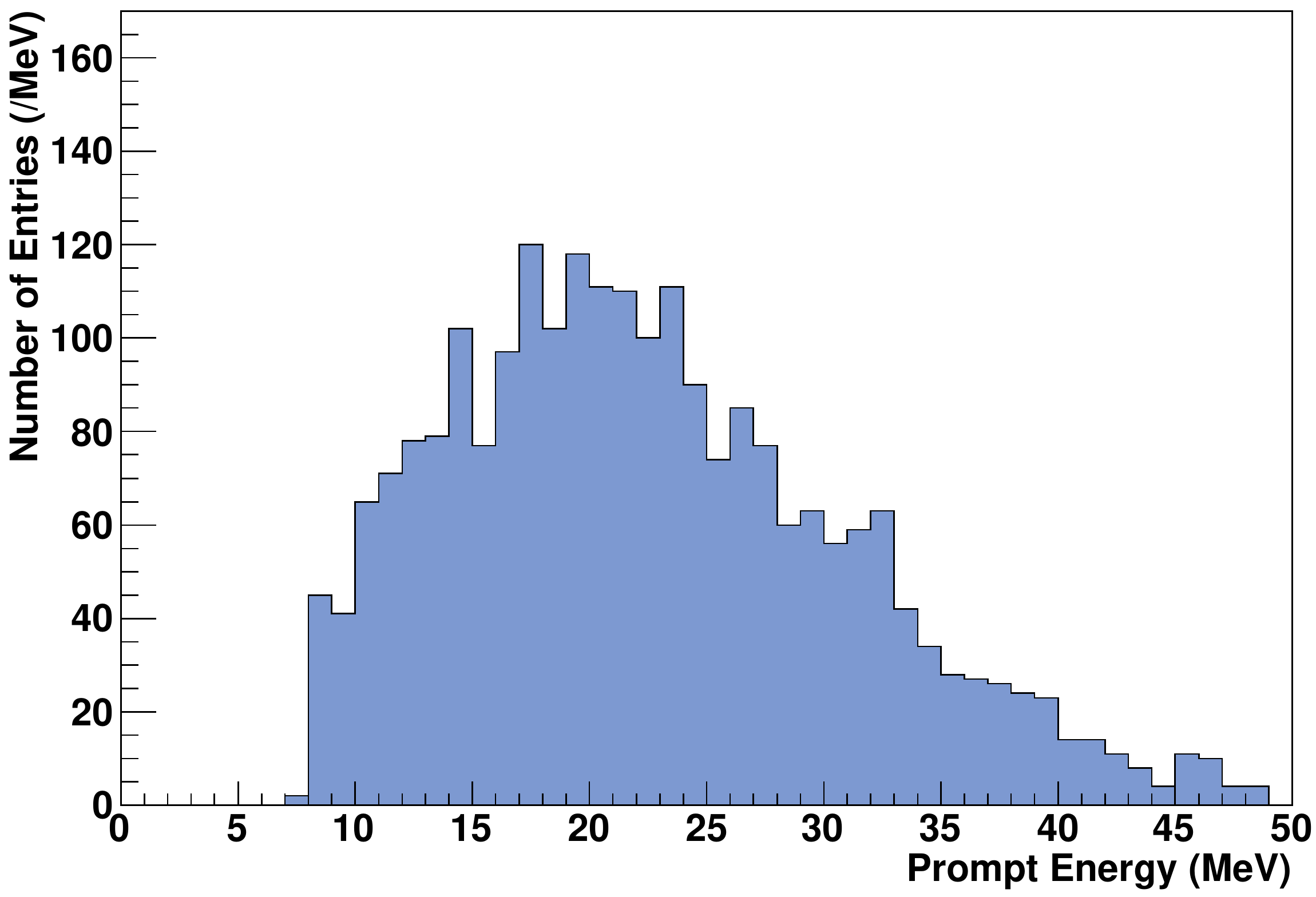}
\caption{The plots are after selection cut with respect to one single AD. Top: Simulation 2-D plot of supernova neutrino selection for delayed signal against prompt signal. Bottom: Prompt signal energy projection of the corresponding 2-D plot above, which indicates the shape of the supernova burst neutrinos. Left: For Gd-LS volume. Right: For LS volume.}
\label{fig:simulation}
\end{figure}

SN$\nu$s have been simulated separately in both the Gd-LS and LS region of a single AD, whose results after selection cuts are shown in Fig.~\ref{fig:simulation}. With these simulation results, the detection efficiency of SN$\nu$ that undergo an IBD in the detector volume is estimated to be $\sim70\%$ and used to determine the expected number of SN$\nu$ events in each AD at Daya Bay.

\section{Supernova online trigger judgment}
An approach is developed to investigate the background coincidence rate, e.g. false alarm rate, by combining all 8 ADs' SN$\nu$ candidate events in the 10s-time-window. The SN$\nu$ candidates are always backgrounds as so rare SN explosion can be observed by neutrino detection in the earth. Every one second, the SN$\nu$ candidates in the previous 10s-time-window are counted in each AD, forming a combination to judge whether to trigger or not.

\subsection{Trigger table and trigger cut}
\label{triggerthreshold}
A trigger table is generated to list the AD background combination cases in order of their corresponding false alarm rate for the sliding 10s-time-window. Utilizing this table, it is convenient to set the cut for the combination cases due to a certain false alarm rate threshold according to SNEWS requirement. Below (Tab.~\ref{tab:triggertable}), part of the trigger table for online test is shown as an example where the contents are all for backgrounds.

In Tab.~\ref{tab:triggertable}, the number under each AD is the background event number counted in the 10s-time-window. The first two columns correspond to the detectors in the Daya Bay near site, the next two columns correspond to the Ling Ao near site and the four remaining columns correspond to the Far site. The column ``False Alarm Rate'' is defines not as the trigger rate relative to the combination in that row but as the total trigger rate of all the AD background combination cases below. Before the false alarm rate calculation, the AD background combination cases are firstly in descending order with respect to their trigger rates. Then for each combination case, the total trigger rate of those below it and itself is calculated serving as the corresponding quantity ``False Alarm Rate''. Obviously, a trigger cut can be determined easily due to the false alarm rate threshold. For example, a 1/34s (0.0293111 Hz) false alarm rate threshold is required and then the last row of Tab.~\ref{tab:triggertable} is where to place the cut below which all the AD background combination cases have a smaller ``False Alarm Rate'' and are supposed to trigger a supernova early warning.

This table is for background false alarm control and SN$\nu$ events are expected to have higher probability for coincidence in 8-ADs than muon-induced fast neutrons or reactor neutrinos, etc. In detailed detection probability for SN explosion, please see Section~\ref{sndetprob}. It is also emphasized here that the ``False Alarm Rate'' in Tab.~\ref{tab:triggertable} is predicted rather than measured. This will be explained in the next subsection.

\begin{center}
\begin{table}[h!]
\centering
\caption{\label{tab:triggertable}Part of the trigger table for supernova online judgment. AD1 to AD8 indicates the 8 antineutrino detectors in the three experimental halls in Daya Bay.}
\begin{tabular}{ccccccccccccc}
\br
AD1 & AD2 & & AD3 & AD4 & & AD5 & AD6 & AD7 & AD8 & & False Alarm Rate (Hz) \\
\mr
0 & 0 & & 0 & 0 & & 0 & 0 & 0 & 0 & & 1 \\
0 & 1 & & 0 & 0 & & 0 & 0 & 0 & 0 & & 0.499092 \\
1 & 0 & & 0 & 0 & & 0 & 0 & 0 & 0 & & 0.404098 \\
\vdots & \vdots & & \vdots & \vdots & & \vdots & \vdots & \vdots & \vdots & & \vdots \\
0 & 0 & & 0 & 1 & & 0 & 1 & 0 & 0 & & 0.0317780 \\
0 & 0 & & 1 & 0 & & 0 & 1 & 0 & 0 & & 0.0305445 \\
0 & 1 & & 0 & 2 & & 0 & 0 & 0 & 0 & & 0.0293111 \\
\vdots & \vdots & & \vdots & \vdots & & \vdots & \vdots & \vdots & \vdots & & \vdots \\
\br
\end{tabular}
\end{table}
\end{center}

\subsection{Background rate prediction}
The reason we use the predicted background rate is that the data sample used for the supernova online trigger study is only about 120 days, which provides insufficient statistics to set a false alarm rate threshold like 1/year. However, the prediction has a challenge -- the overlap in the sliding 10s-time-window -- every one second, the SN$\nu$ candidates of each AD are combined for judgment and the 10s-time-window is overlapped by a few adjacent ones.

For a single AD, it is verified with the numerical simulation that the rate (Hz) (here the probability is numerically equal to the rate as every one second there is a combination) of the event count in the sliding 10s-time-window still follows the Poisson distribution with the mean value $10~seconds~\times~single~AD~event~rate$. This is the fundament of the combination calculation.

In terms of the combination of multiple ADs, assuming different experimental halls are mutually independent for backgrounds, the correlation between ADs in the same site is considered and measured using the data. The correlation between ADs in the same site originates from the muon-induced fast neutrons which cause several consecutive signals in detectors of the same experiment hall. The trigger rate for each combination case is predicted using several unknown independent Poisson variables that formulate the event rate of each AD and some of which are shared by the correlated ADs in the same experimental halls representing the correlation part. These unknown Poisson variables can be calculated eventually based on the measured single AD event rates and correlation between ADs.

%For instance, in the case that 2-AD in one experimental hall, three independent Poisson variables are defined: $X_1, X_2, X$. Suppose AD1 event count is $Y_1=X_1+X$ and AD2 event count is $Y_2=X_2+X$, the mean values for the three variables are able to be deduced from the measured covariance $Cov(Y_1,Y_2)=X$ and the measured mean values of $Y_1, Y_2$. Therefore, each combination case $(Y_1, Y_2)$ could be decomposed into the possible cases $(X_1, X_2, X)$ and the corresponding probability is easily obtained.

In addition, given the trigger rate of each combination case, the statistical error can be derived utilizing some statistic skills in which case the data sample has to be split into 10 parts according to the time and each of the 10 parts is 1s delay or earlier than the adjacent one. To verify the prediction, the rates measured on data are compared to the prediction and 82\% are within 1$\sigma$, 98.4\% are within 2$\sigma$, and 99.7\% are within 3$\sigma$ consistent with the prediction. Therefore, the prediction of background combination rate is plausible to replace the measured one. Notice that the systematic error is negligible compared with the statistical error when the threshold is set too small such as 1/year, or even 1/month.

\subsection{Supernova online trigger diagram}
\begin{figure}[h!]
\centering
\includegraphics[width=8cm]{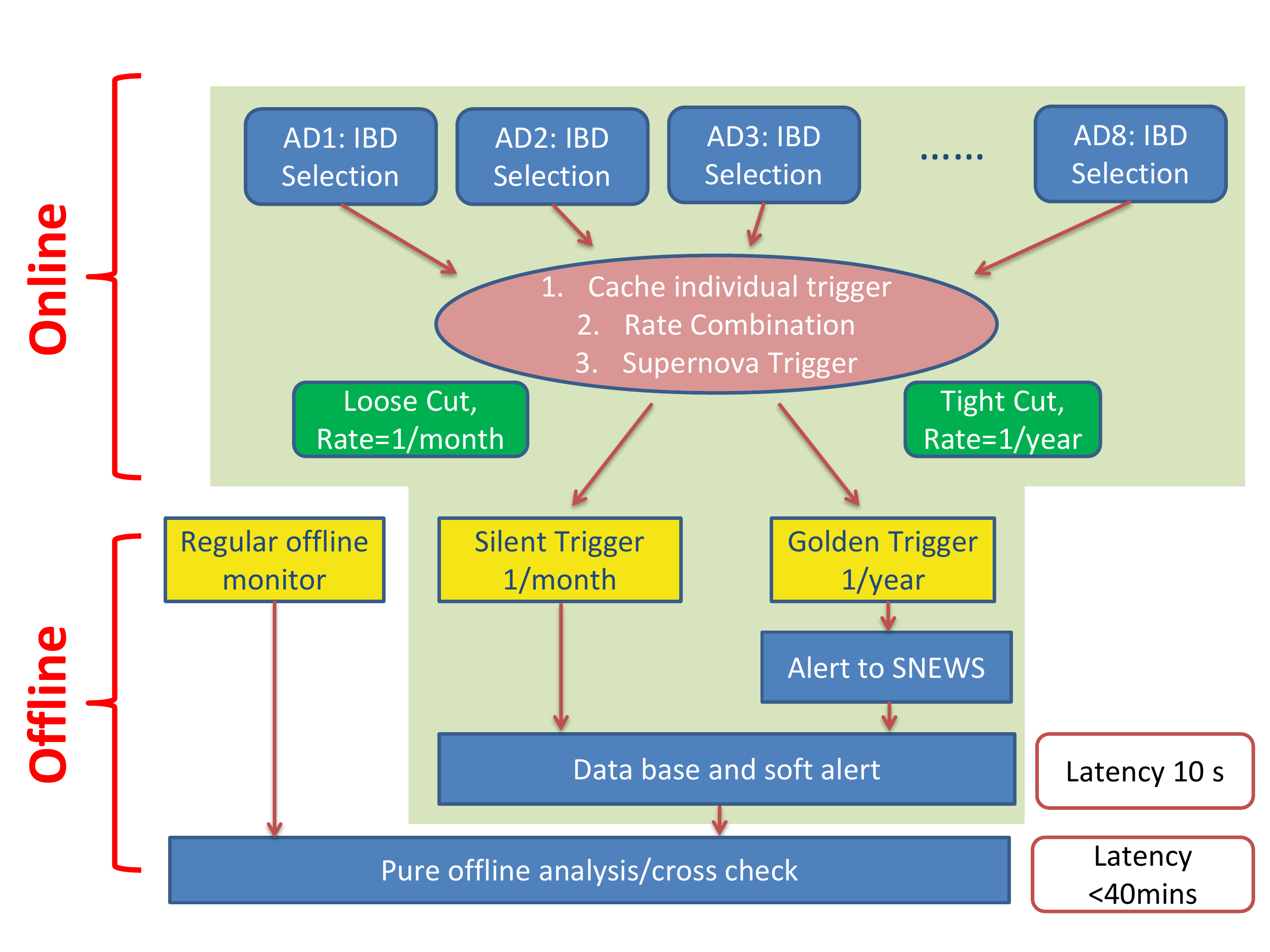}
\caption{Diagram of supernova online trigger system in Daya Bay. It is the framework of the software applications on the basis of the existing DAQ system and on-site host.}
\label{fig:snews_diagram}
\end{figure}

The scheme of the supernova online trigger system in Daya Bay is shown in Fig.~\ref{fig:snews_diagram}. It includes several software applications implemented in the DAQ of Daya Bay.

The online part is able to get access to all the raw data and make a simple reconstruction. The IBD selection program for each AD provides the information of SN$\nu$ candidates to a combination server with the function of combination and trigger judgment according to the trigger table mentioned above. There are two levels of trigger, silent trigger (1/month) and golden trigger (1/year), which are related to different offline responses. In case of a golden trigger, an e-mail alert is immediately sent and information of those SN$\nu$ candidates is written into an offline database with about 10 seconds time latency. A pure offline analysis would cross check both the golden and silent triggers with less than 40 min latency. The shaded area in the diagram has been tested and officially installed, while the offline analysis/cross-check is being developed based on the Performance Quality Monitoring System (PQM) of Daya Bay. Daya Bay is negotiating to join the SNEWS and the e-mail alert is presently sent to Daya Bay collaborators who are interested.

To exclude unexpected trigger bursts (e.g.~electronic noise) in one detector or one experimental hall, a simple but effective uniformity cut based on the $\chi^2$ method is applied with less than 1\% detection probability lost for supernova explosions. This $\chi^2$ is the minimum value of $\sum_i\frac{(n_i-\lambda)^2}{n_i}$ where $n_i$ is the event counts in the combination for each AD and $\lambda$ is the best fit value of event counts for all ADs considering SN$\nu$ events are distributed uniformly among ADs. Detection probability of a supernova explosion is explained next section.

\section{Detection probability of a supernova explosion}
\label{sndetprob}
According to the target mass of the Daya Bay detectors, the detection efficiency of SN$\nu$ obtained based on MC and the relation between supernova neutrino time-integrated flux and distance to the earth \cite{snevent}, single AD's expected SN$\nu$ event counts can be determined below,
$$
N_{AD} = N_0\times\frac{L_{\bar{\nu}_e}}{5\times10^{52}erg}\times(\frac{10kpc}{D})^2
$$
where $L_{\bar{\nu}_e}$ is the luminosity of electron-antineutrino emission and $D$ is the SN explosion distance to earth. $N_0$ is the single AD's expected SN$\nu$ event number in 10s-time-window corresponding to $5\times10^{52}erg$ luminosity and $10~kpc$ distance. Detection efficiency of SN$\nu$ is considered in $N_0$.

\begin{figure}[h!]
\centering
\includegraphics[width=8cm]{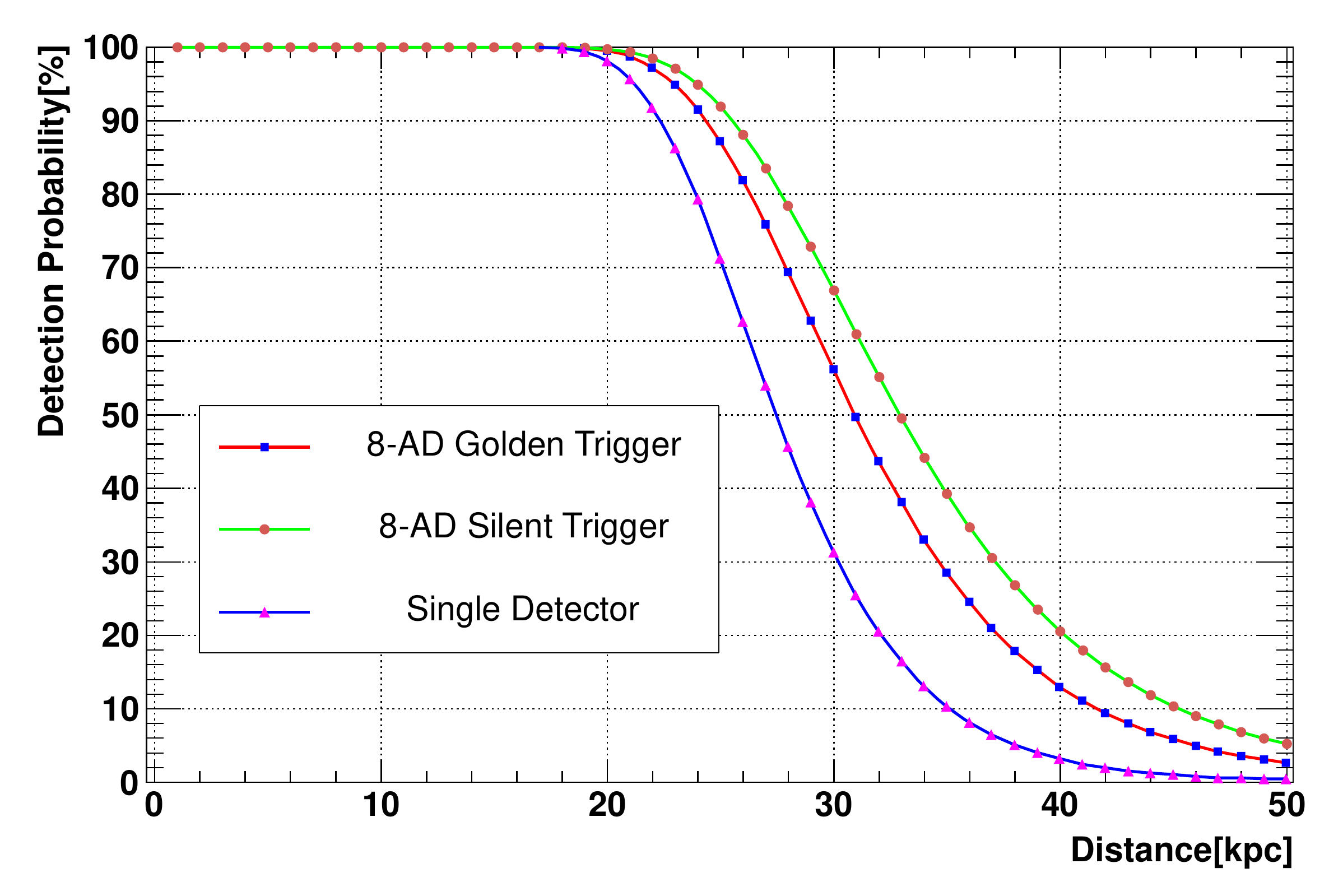}
\caption{The X-axis is SN explosion distance from the earth and the Y-axis is the corresponding detection probability. ``8-AD golden trigger'' corresponds to the result with false alarm rate $<$1/year, and ``8-AD silent trigger'' corresponds to that with false alarm rate $<$1/month. ``Single Detector'' is the scenario also with false alarm rate $<$1/year in which the 8-AD target mass is combined into a single detector with the background rate level per target mass of Daya Bay ADs.}
\label{fig:detectionprobability}
\end{figure}

Here, the supernova model for the detection probability calculation is set to SN1987A-type and a typical value for $N_{AD}$ is $\sim$8 at a distance of 10 kpc to earth. Based on the expected SN$\nu$ events of each AD, the detection probability of a supernova explosion is calculated by summing up the probabilities of the combination cases that pass the trigger threshold. Notice that the single AD event rate increases simultaneously during a supernova explosion and coincidence signals in multiple ADs occur more frequently. As a result, the detection probability of the SN explosion has been calculated as a function of distance to the earth. The result is shown in Fig.~\ref{fig:detectionprobability}. From the ``8-AD Golden Trigger'' line, the Milky Way center is around 8.5 kpc from the earth with a 100\% detection probability and the most distant edge of the Milky Way is 23.5 kpc from the earth with a 94\% detection probability. Moreover, the silent trigger will add a potential 5\% to 10\% detection probability of SN explosion.

Particularly, the ``Single Detector'' line is comparable to the ``8-AD Golden Trigger'' which obviously indicates the gain in sensitivity of the 8-AD configuration over a single detector. A rough estimation implies the Daya Bay is equivalent to a single 0.7 kton liquid scintillator detector with respect to the detection probability of SN explosion as a consequence of the multi-AD configuration. The background rate level per target mass is the average background rate per unit of the target mass of Daya Bay ADs.

The detection probability has two elements in reality: one is the probability for a SN explosion can be detected, the other one is the corresponding ``false alarm rate'' threshold (defined in Subsection~\ref{triggerthreshold}), for example, 1/month or 1/year here which is for background false alarm control. Based on this, the difference between single-detector and multi-detector can be explained. In the scenario of single-detector, the total number of SN$\nu$ events is exploited for trigger cut setting, for example, 10 SN$\nu$ events in 10s-time-window corresponding to 1/month false alarm rate threshold. While in the scenario of multi-detector, the background combination case is exploited for trigger cut setting, for example, combination 0-0-2-3-1-1-0-0 corresponding to 1/month false alarm rate threshold. Obviously, the total number of events in multi-detector here is 7 which is smaller than the single-detector, thus providing a higher detection probability.

\section{Summary}
The supernova online trigger system in Daya Bay has been officially installed after several pretests. The extra workload to the current CPU consumption of DAQ is around 8\% and is far from the computing maximum workload online. Moreover, the time latency from electronics triggers to an alarm is around 10 s (20 s considering the duration of 10s-time-window). In the future, the pure offline cross check will be added and joining the SNEWS is underway. With a relatively low energy threshold, superior energy resolution and separated 8-AD deployment, the online detection probability for a SN1987A-type SN explosion could be larger than 94\% within the Milky Way.

\ack
This work is supported in part by the Ministry of Science and Technology of China and the National Natural Science Foundation of China (Grants No.11235006). In addition, the author also wishes to acknowledge the Daya Bay Reactor Neutrino Experiment Collaboration, particularly Shaomin Chen, Zhe Wang, Logan Lebanowski and Fei Li for precious information, useful discussion and selfless help.

\end{document}